\documentclass[journal]{IEEEtran}

\usepackage{myStyleIEEE}
\usepackage{nomencl}
\usepackage{ragged2e}
\IEEEoverridecommandlockouts

\usepackage{etoolbox}
\renewcommand\nomgroup[1]{%
  \item[\bfseries
  \ifstrequal{#1}{P}{A. Parameters}{%
  \ifstrequal{#1}{V}{C. Variables}{%
  \ifstrequal{#1}{S}{B. Sets and Indices}{}}}%
]}

\begin{document}
\bstctlcite{IEEEexample:BSTcontrol}

\title{Voltage Stability Constrained Unit Commitment in High IBG-Penetrated Power Systems }

\newtheorem{proposition}{Proposition}
\renewcommand{\theenumi}{\alph{enumi}}

\author{Zhongda~Chu,~\IEEEmembership{Student~Member,~IEEE,} 
        Fei~Teng,~\IEEEmembership{Senior Member,~IEEE} 
        
        
\vspace{-0.5cm}}
\maketitle
\IEEEpeerreviewmaketitle

\begin{abstract}
With the increasing penetration of renewable energy sources, power system operation has to be adapted to ensure the system stability and security while considering the distinguished feature of the Power Electronics (PE) interfaced generators. The static voltage stability which is mainly compromised by heavy loading conditions in conventional power systems, faces new challenges due to the large scale integration of PE-interfaced devices. This paper investigates the static voltage stability problem in high PE-penetrated system. The analytic criterion that ensures the voltage stability at the Inverter-Based Generator (IBG) buses are derived with the interaction of different IBGs being considered. Based on this, an optimal system scheduling model is proposed to minimize the overall system operation cost while maintaining the voltage stability during normal operation through dynamically optimizing the active and reactive power output from IBGs. The highly nonlinear voltage stability constraints are effectively converted into Second-Order-Cone (SOC) form, leading to an overall Mixed-Integer SOC Programming (MISOCP), together with the SOC reformulation of AC power flow and frequency constraints. The effectiveness of the proposed model and the impact of various factors on voltage stability are demonstrated in thorough case studies.
\end{abstract}


\makenomenclature
\mbox{}
\nomenclature[P]{$J$}{Lumped inertia of WT driven systems$\,[\mathrm{Mkgm}^2]$}

\section{Introduction} \label{sec:1}
A transition of traditional Synchronous Generator (SG) based power systems to incorporate fast growing renewable energy, such as wind and solar, has been witnessed in the last few decades due to the environmental concerns and related policies. Unlike the conventional SGs with well-defined active and reactive power control and inertia provision, Inverter-Based Generators (IBGs) present completely different features. They have intermittent power outputs depending on the available Renewable Energy Resources (RESs) and do not inherently provide active/reactive power support. In addition,
the low short-circuit capacity of IBGs worsen the system performance during fault conditions and leads to buses with low system strengths \cite{electronics10020115,9329077}. 

These distinguished characteristics of IBGs bring variations and challenges to power system operation and stability \cite{8450880,Hatziargyriou2020}. In particular, voltage stability problems in high PE penetrated systems have been reported \cite{NERC,AEMO} and attract significant attentions from researchers \cite{MODARRESI20161,Ar2021}. Conventionally, voltage stability analysis is typically carried out at load buses while assuming that the voltages at SG buses can be well regulated \cite{Cutsem1998}. Hence, the voltage stability is mainly influenced by the loading conditions in the system. However, with the increasing integration of grid-following IBGs, the voltage stability at IBG buses needs to be reassessed. The voltages at IBG buses are not directly controlled by the converters and are the results of active and reactive power regulations. In addition, most of RESs are located far away from the load center, leading to large electrical distances and thus low system strengths, which may result in the voltage instability at IBG buses and other more severe events. 

Voltage stability analysis has been recently carried out in power systems with high PE penetration. In \cite{8486639}, different voltage stability criteria in a single-infeed Voltage-Source Converter (VSC) HVDC system are derived based on the voltage sensitivity factor and it is shown that the power-voltage sensitivity characteristics of VSC-HVDC system can be improved by modulated reactive power control. A grid-feeding converter voltage stability assessment method is proposed in \cite{8728057}. Based on the power-voltage curve, the maximum deliverable power from the converter to the grid, beyond which the converter becomes voltage unstable is found. The voltage stability of IBGs under weak-grid conditions is analyzed in \cite{8769713} and \cite{ALSOKHIRY2020105899} where the loss of voltage stability stems from excessive active power export and countermeasures are proposed to improve the active-power transfer capacity. \cite{WANG2021107087} investigates the impact of non-synchronous machine sources on small and large disturbance voltage stability based Jacobian matrix singularity and transient process analysis. 

Nevertheless, these studies are limited to simple systems with single or few converter-based units and the interactions among converters and the rest of the system are not considered. The voltage stability in a generic distribution system is assessed in \cite{8574925} through network-load admittance ratio, which is proven to be equivalent to power flow Jacobian singularity. The value of using time-series ac power flow analysis techniques in assessing voltage stability is demonstrated in \cite{5337959} and the voltage stability margin of the power system is increased by the implementation of voltage control strategies in wind turbines. To account for the interactions among electrically interconnected renewable energy sources in weak power systems, a site-dependent short-circuit ratio is proposed in \cite{8074765}, revealing the relationship between system strength and static voltage stability but the impact of reactive power support on voltage stability is neglected.

On the other hand, it is essential to maintain the voltage stability in power systems with high PE penetration and there are few researches considering the voltage stability during the system scheduling process. A voltage stability-constrained optimal power-flow model is proposed in \cite{8279490} where the voltage stability constraints are derived based on the power-flow Jacobian nonsingularity and formulated in second-order cone form. However, the impacts of IBGs on voltage stability are not considered. \cite{FURUKAKOI2018618} proposes a multipurpose operation planning method for minimizing the prediction error of photovoltaic power and improving voltage stability of power systems, where the RESs and the impact of generator status on the voltage stability are not considered. A Unit Commitment (UC) problem with voltage stability constraints is presented in \cite{2020JEET,app9163412} where sufficient voltage supporting capacity from thermal plants at HVDC feed-in area is scheduled to ensure the voltage stability. However, the highly nonlinear constraints and the complex formulation requires iteration between the master UC problem and voltage stability sub-problems, which increases the model complexity. Short-term voltage stability constraints are considered in unit commitment problem and microgrid scheduling by \cite{9224156,wrro175989} respectively, where dynamic simulation is embedded in the optimization problem to quantify the voltage stability. 

In this context, this work proposes a voltage stability constrained UC problem in high PE penetrated systems to ensure the static voltage stability of IBGs while considering their interactions. The main contributions of this paper are summarized as follows:
\begin{itemize}
    \item The static voltage stability constraints are derived in a general high PE-penetrated system, where the interaction among IBGs are quantified through the system impedance matrix. The highly nonlinear voltage stability constraints involving matrix inverse are further effectively reformulated into SOC form.
    \item An optimal system scheduling model is proposed to minimize the overall system operation cost while maintaining the static voltage stability of IBGs during normal operation through dynamically optimizing the active and reactive power injection from IBGs. Together with the SOC relaxation of AC power flow and frequency constraints, the overall problem is formulated as MISOCP.
    \item The effectiveness of the proposed model and the impacts of maintaining voltage stability on system operation under different control strategies are investigated through case studies based on IEEE 30-bus system.
\end{itemize}

The remainder of this paper is organized as follows. Section \ref{sec:2} analyzes the static voltage stability at IBG buses in a general high PE-penetrated system and the voltage stability constraints are derived. The resulted highly nonlinear constraints are further reformulated into SOC form in Section \ref{sec:3}. The overall voltage stability constrained UC problem is introduced in Section \ref{sec:4}. Case studies are carried out in Section \ref{sec:5} and Section \ref{sec:6} concludes the paper.

\section{Voltage Stability Analysis} \label{sec:2}
Consider a power system having $n\in\mathcal{N}$ buses with $g_c\in\mathcal{G}_c$, $g_v\in\mathcal{G}_v$ and $c\in\mathcal{C}$ being the set of conventional Synchronous Generators (SGs), grid-forming Virtual Synchronous Generators (VSGs) and grid-following Inverter-Based Generators (IBGs). $\Psi(g)$ and $\Phi(c)$ map the units in $g\in \mathcal{G}=\mathcal{G}_c\bigcup\mathcal{G}_v$ and $c\in \mathcal{C}$ to the corresponding bus indices respectively. In this work, static voltage stability refers to the ability of power systems to maintain acceptable steady-state voltages at all buses in the system under normal operating conditions and following small disturbances. Note that the grid-forming VSGs are treated the same as the SGs since their terminal voltages can be well regulated during steady state and the power saturation would not occur following small disturbances. The critical condition of voltage stability can be derived based on different methods, namely the power-voltage (PV) curve, the voltage to power sensitivity ($d|V|/dP$) and the singularity of the power flow Jacobian matrix. Since they are mathematically equivalent and the same result can be obtained, only the last one is covered here for voltage stability derivation.
\begin{figure}[!t]
    \centering
    \vspace{-0.4cm}
	\scalebox{0.65}{\includegraphics[]{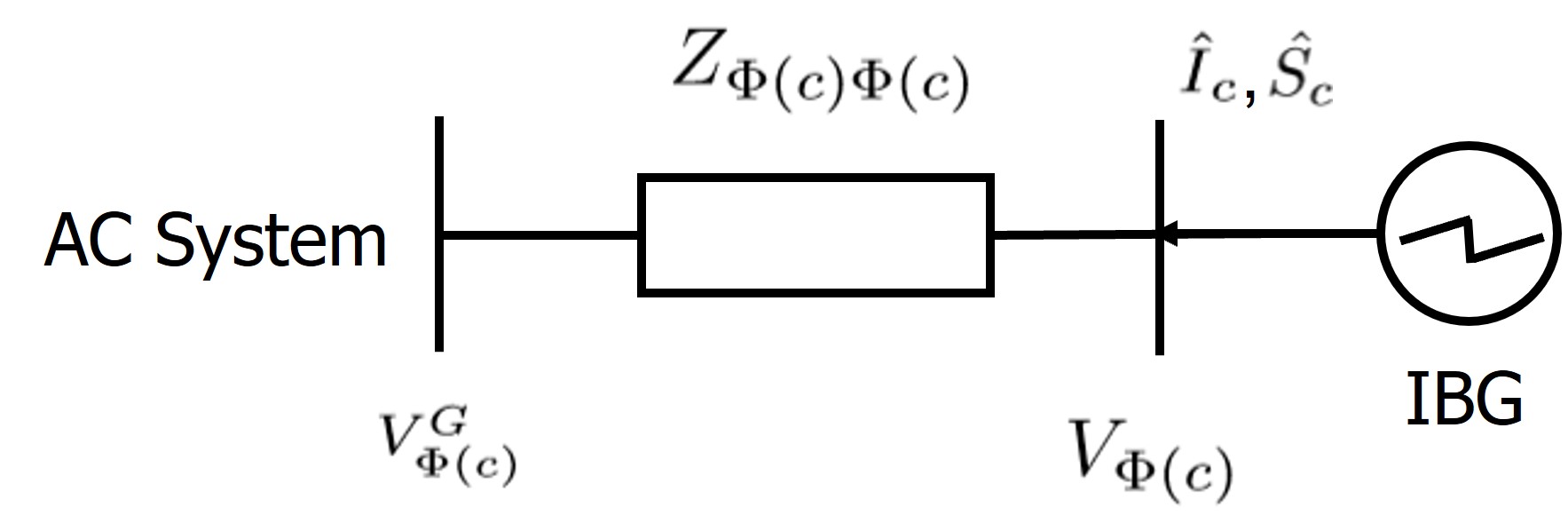}}
    \caption{\label{fig:2bus}Equivalent circuit of a general AC system seen from an IBG bus.}
\end{figure}

\begin{figure}[!t]
    \centering
	\scalebox{0.52}{\includegraphics[]{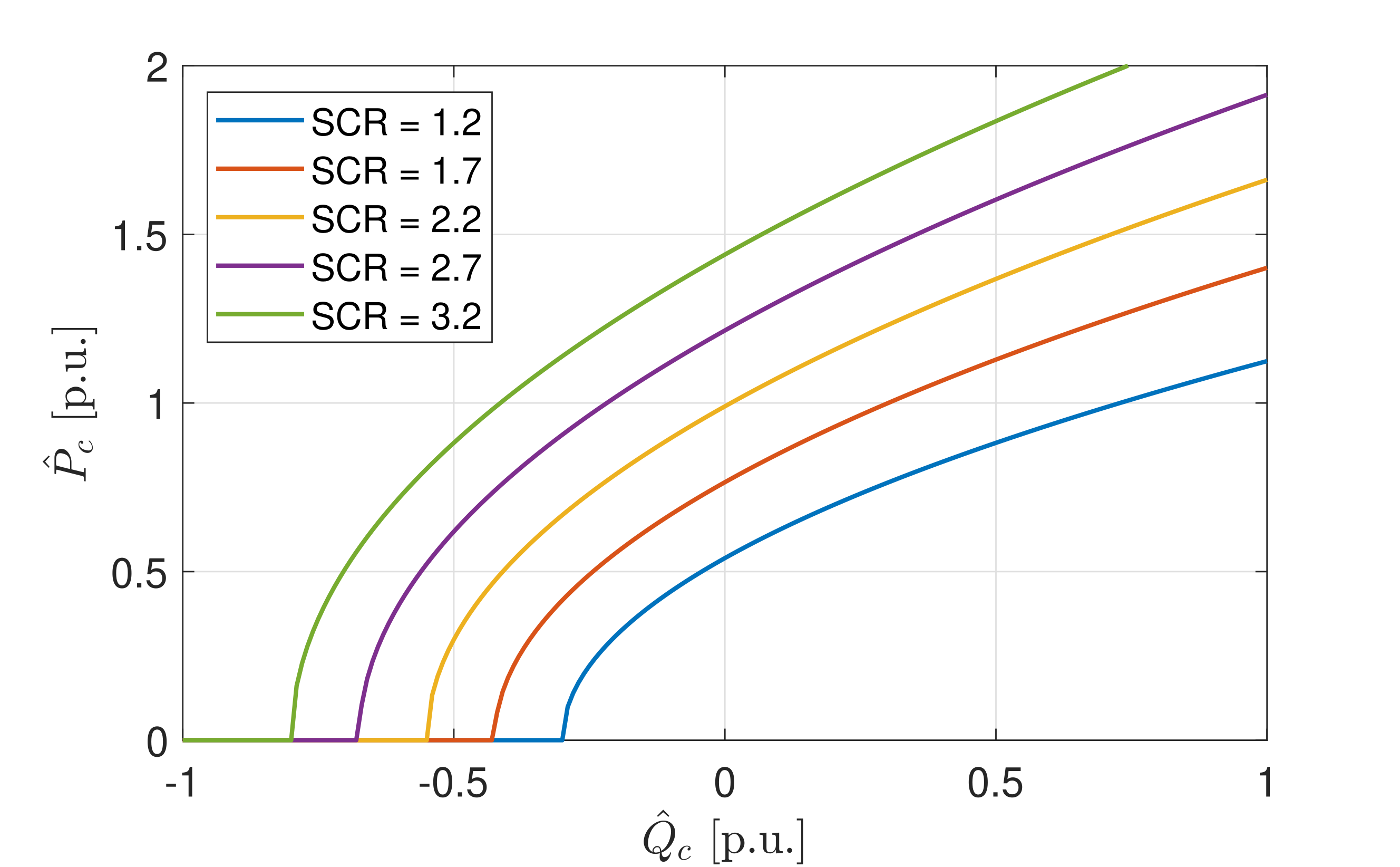}}
    \caption{\label{fig:PQ_Curve}Boundaries of voltage stability with different SCR.}
\end{figure}

At steady-state, the system power flow equations can be represented in the following matrix form:
\begin{equation}
\label{PF}
    V = Z I
\end{equation}
where $V$ and $I$ are the vector of voltages and current injections at all the buses and $Z$ is the system impedance matrix. Specifically, the voltage at the IBG bus can be obtained by extracting the $\Phi(c)^{\mathrm{th}}$ row of \eqref{PF}:
\begin{equation}
    \label{V_c}
    V_{\Phi(c)} = \sum_{g\in\mathcal{G}}Z_{\Phi(c)\Psi(g)} I_{g}+\sum_{c'\in\mathcal{C}}Z_{\Phi(c)\Phi(c')}I_{c'}
\end{equation}
with $I_g$ and $I_{c'}$ being the current injection from (V)SG $g$ and IBG $c'$. Based on the formulation in \cite{8074765}, the relationship in \eqref{V_c} can be represented by an equivalent two-bus system with one bus being Bus $\Phi(c)$ and the other bus representing the rest of the system, i.e., the equivalent grid bus with its voltage denoted by $V^G_{\Phi(c)}$:
\begin{align}
\label{V_c_eq}
    V_{\Phi(c)} & = \underbrace{\sum_{g\in\mathcal{G}}Z_{\Phi(c)\Psi(g)} I_{g}}_{V^G_{\Phi(c)}} \nonumber \\
    & + Z_{\Phi(c)\Phi(c)} \underbrace{\left(I_c + \sum_{c'\in\mathcal{C}, c'\neq c}\frac{Z_{\Phi(c)\Phi(c')}}{Z_{\Phi(c)\Phi(c)}}I_{c'}\right)}_{\hat I_{c}}.
\end{align}
As shown in Fig.~\ref{fig:2bus}, the IBG bus is connected to the grid bus through the impedance $Z_{\Phi(c)\Phi(c)}$ with an equivalent current injection, $\hat{I}_c$. This equivalent current includes the current from IBG $c$ and the currents from other IBGs referred to Bus $\Phi(c)$ by the impedance ratio. As a result, the equivalent complex power injection from IBG c is calculated by:
\begin{align}
\label{S_c1}
    \hat{S_{c}}&=V_{\Phi(c)}\hat I_{c}^*.
\end{align}
Combine \eqref{V_c_eq} and \eqref{S_c1} and neglect the resistance in $Z$ matrix due to the small $R/X$ ratio in transmission system leading to:
\begin{align}
\label{S_c2}
    \hat{S_{c}}&=V_{\Phi(c)} {\left( \frac{V_{\Phi(c)}-V^G_{\Phi(c)}}{Z_{\Phi(c)\Phi(c)}}\right)}^* \nonumber \\
    &= \underbrace{ \frac{|V_{\Phi(c)}||V^G_{\Phi(c)}|\sin \theta_c}{-|Z_{\Phi(c)\Phi(c)}|}}_{\hat{P}_c}+\mathrm{j}\underbrace{\frac{|V_{\Phi(c)}|^2-|V_{\Phi(c)}||V^G_{\Phi(c)}|\cos \theta_c}{|Z_{\Phi(c)\Phi(c)}|}}_{\hat{Q}_c}
\end{align}
where $\theta_c = \angle{V^G_{\Phi(c)}} -\angle{V_{\Phi(c)}}$ is the angle difference between the equivalent grid bus and the IBG bus; $\hat P_c$ and $\hat Q_c$ are the equivalent active and reactive power injection from IBG $c$. Note that the negative sign in $\hat P_c$ is due to the direction of $\hat I_c$ defined in \eqref{V_c_eq}. The Jacobian matrix $J$ of $\left[\hat P_c(|V_{\Phi(c)}|,\theta_c),\,\hat{Q}_c(|V_{\Phi(c)}|,\theta_c)\right]^{\mathrm{T}}$ can be obtained as follows:
\begin{align}
    J =& \frac{1}{|Z_{\Phi(c)\Phi(c)}|} \cdot \nonumber \\
    &\begin{pmatrix} -|V^G_{\Phi(c)}|\sin \theta_c & -|V^G_{\Phi(c)}||V_{\Phi(c)}|\cos \theta_c\\ 2|V_{\Phi(c)}|-|V^G_{\Phi(c)}|\cos \theta_c& |V^G_{\Phi(c)}||V_{\Phi(c)}|\sin \theta_c \end{pmatrix}
\end{align}
The singularity of the Jacobian matrix would lead to  voltage instability at Bus $\Phi(c)$. Therefore, solving $|J|=0$ and eliminating the angle $\theta_c$ with $\hat P_c$ and $\hat Q_c$ as defined in \eqref{S_c2} gives the following condition of voltage stability:
\begin{equation}
\label{VS}
    \hat{P}_c^2+\hat{Q}_c^2\le { \Bigg( \hat{Q}_c+ \underbrace{ \frac{{|V^G_{\Phi(c)}}|^2}{2|Z_{\Phi(c)\Phi(c)}|}}_{\Gamma_c} \Bigg) ^2}.
\end{equation}
This voltage stability condition illustrates that the equivalent active power injection of IBG c is limited by an upper bound depending on short circuit capacity (${{|V^G_{\Phi(c)}}|^2}/{|Z_{\Phi(c)\Phi(c)}|}$) at Bus $\Phi(c)$ if no reactive power is supplied. In weak points of the system where the short circuit capacity is low, the equivalent active power injection may have to be set smaller than the rated or even the available power to maintain voltage stability. Furthermore, \eqref{VS} also indicates that properly setting the equivalent reactive power would allow more active power to be delivered. This can be justified by taking the derivative $\hat{P}_c$ with respect to $\hat{Q}_c$ on the boundary of \eqref{VS}:
\begin{equation}
    \frac{\partial \hat P_c}{\partial \hat Q} = \frac{1}{2}\left( \frac{QV^2}{Z}+\frac{V^4}{4Z} \right)^{-\frac{1}{2}}\cdot \frac{V^2}{Z}>0.
\end{equation}
An example of the static voltage stability boundaries at IBG bus c with various Short Circuit Ratios (SCR) is shown in Fig.~\ref{fig:PQ_Curve}. The SCR in per unit is defined as:
\begin{equation}
    \mathrm{SCR} = \frac{{|V^G_{\Phi(c)}}|^2}{|Z_{\Phi(c)\Phi(c)}|\cdot \mathbf{P}_c},
\end{equation}
where $\mathbf{P}_c$ is the capacity of IBG c. It can be observed from the figure that the maximum equivalent active power transfer capability is reduced as the system strength (SCR) becomes lower and increasing the IBG reactive power injection facilitates the active power transfer. 

It should be noted that the voltage stability at Bus $\Phi(c)$ does not only depend on the power injections from IBG c but also that from other IBGs. In order to better reveal this interaction of different IBGs, $\hat P_c$ and $\hat Q_c$ are explicitly expressed by inserting $\hat{I}_c$ defined in \eqref{V_c_eq} into \eqref{S_c1} as follows:
\begin{align}
    \hat{S_{c}}& =   S_{c}+\sum_{c'\in \mathcal{C},c'\neq c}\frac{Z_{\Phi(c)\Phi(c')}}{Z_{\Phi(c)\Phi(c)}} \frac{V_{\Phi(c)}}{V_{\Phi(c')}}S_{c'} \nonumber \\
    &= \underbrace{\left(P_{c}+  \Re\left(  \sum_{c'\in \mathcal{C},c'\neq c}\frac{Z_{\Phi(c)\Phi(c')}}{Z_{\Phi(c)\Phi(c)}} \frac{V_{\Phi(c)}}{V_{\Phi(c')}}S_{c'} \right) \right)}_{\hat P_c} \nonumber \\
    & +\mathrm{j} \underbrace{\left(Q_{c}+ \Im\left( \sum_{c'\in \mathcal{C},c'\neq c}\frac{Z_{\Phi(c)\Phi(c')}}{Z_{\Phi(c)\Phi(c)}} \frac{V_{\Phi(c)}}{V_{\Phi(c')}}S_{c'}\right) \right)}_{\hat{Q}_c}.
\end{align}
The equivalent power injections from IBG c ($\hat P_c/\hat Q_c$) accounts for the local power ($P_c/Q_c$) and the power from other IBGs, which is referred to the concerned bus according to the ratio of $Z$ matrix elements and the bus voltages. Since the relationship $\frac{|Z_{\Phi(c)\Phi(c')}|}{|Z_{\Phi(c)\Phi(c)}|}<1,\,\frac{V_{\Phi(c)}}{V_{\Phi(c')}}\approx1$ holds in a general system during normal operation, the range of $\hat P_c \in [0,2]\, \mathrm{p.u.}$ is selected in Fig.~\ref{fig:PQ_Curve}.

\section{SOC Reformulation of Voltage Stability Constraints} \label{sec:3}
Although the voltage stability constraint described in \eqref{VS} is in SOC form in terms of $\hat P_c$, $\hat Q_c$ and $\hat Q_c+ \Gamma_c$, the expression of $\hat P_c$, $\hat Q_c$, $V^G_{\Phi(c)}$ and $Z_{\Phi(c)\Phi(c)}$ themselves are highly nonlinear with respect to the decision variables. To deal with the nonlinearity in $\hat P_c$ and $\hat Q_c$, the following assumptions are made: (i) the resistance in $Z$ matrix can be neglected due to the small $R/X$ ratio in transmission system; (ii) the voltages at each IBG bus are close to each other under normal operating conditions, i.e., ${V_{\Phi(c)}/}{V_{\Phi(c')}}\approx 1$ \cite{WU201972}. As a result, $\hat P_c$ and $\hat Q_c$ can be simplified as follows:
\begin{subequations}
\label{PQ_1}
\begin{align}
    \hat P_c &= P_{c}+   \sum_{c'\in \mathcal{C},c'\neq c}\frac{|Z_{\Phi(c)\Phi(c')}|}{|Z_{\Phi(c)\Phi(c)}|}P_{c'} \\
    \hat Q_c &= Q_{c}+   \sum_{c'\in \mathcal{C},c'\neq c}\frac{|Z_{\Phi(c)\Phi(c')}|}{|Z_{\Phi(c)\Phi(c)}|}Q_{c'}.
\end{align}
\end{subequations}
Similarly, assume $|V^G_{\Phi(c)}|^2\approx1$ in $\Gamma_c$ giving:
\begin{equation}
    \label{Gamma_c}
    \Gamma_c = \frac{1}{2|Z_{\Phi(c)\Phi(c)}|}.
\end{equation}
However, \eqref{PQ_1} and \eqref{Gamma_c} are still nonlinear equality constraints in a UC problem, since the elements in $Z$ matrix are affected by the (V)SG operating conditions. As a result, the only nonlinearity within the SOC expression in \eqref{VS} is related to the elements in $Z$ matrix, i.e., ${|Z_{\Phi(c)\Phi(c')}|}/{|Z_{\Phi(c)\Phi(c)}|}$ in $\hat P_c$ and $\hat Q_c$ and $1/{|Z_{\Phi(c)\Phi(c)}|}$ in $\Gamma_c$. 

\subsection{Z Matrix Formulation and Linearization}
By definition, the Z matrix can be obtained by taking the inverse of system admittance matrix $Y$ as follows:
\begin{subequations}
\label{Y}
    \begin{align}
    Z &= Y^{-1}\\
    Y &= Y^0 +  Y^g,
    \end{align}
\end{subequations}
where $Y^0$ is the admittance matrix of the transmission lines only; $Y^g$ denotes the additional $Y$ matrix increment due to (V)SGs' reactance. Depending on operating conditions of the (V)SGs, the elements in $Y^g$ can be expressed as:
\begin{equation}
\label{Y2}
    Y_{ij}^g=
    \begin{cases}
    \frac{1}{X_{g_c}}x_{g_c}\;\;&\mathrm{if}\,i = j \land \exists\, g_c\in \mathcal{G}_c,\, \mathrm{s.t.}\,i=\Psi(g_c)\\
    \frac{1}{X_{g_v}}\alpha_{g_v}\;\;&\mathrm{if}\,i = j \land \exists\, g_v\in \mathcal{G}_v,\, \mathrm{s.t.}\,i=\Psi(g_v)\\
    0\;\;& \mathrm{otherwise}.
    \end{cases}
\end{equation}
It should be noted that $x_{g_c},\,\forall g_c \in \mathcal{G}_c$ can be viewed as binary decision variables, whereas for the VSGs, their operating conditions are determined by the current available wind/solar resources rather than the system operator. Therefore, a scenario-dependent parameter $\alpha_{g_v}\in [0,1],\,\forall g_v \in \mathcal{G}_v$ is introduced to represent the percentage of VSGs' online capacity. The approach proposed in \cite{6672214} and \cite{7370811} is used in this paper where the online wind capacity is estimated given the current available power based on historical data.

Although it is theoretically possible to derive each element in $Z$ as a function of $x_{g_c}$ and $\alpha_{g_v}$ according to \eqref{Y} and \eqref{Y2}, the expression becomes extremely complicated as the dimension of $Z$ increases, which cannot be included in the UC problem. In order to effectively linearize ${|Z_{\Phi(c)\Phi(c')}|}/{|Z_{\Phi(c)\Phi(c)}|}$ and $1/{|Z_{\Phi(c)\Phi(c)}|}$ in \eqref{PQ_1} and \eqref{Gamma_c}, a linear regression method is applied. 

\begin{figure}[!t]
    \centering
	\scalebox{0.43}{\includegraphics[]{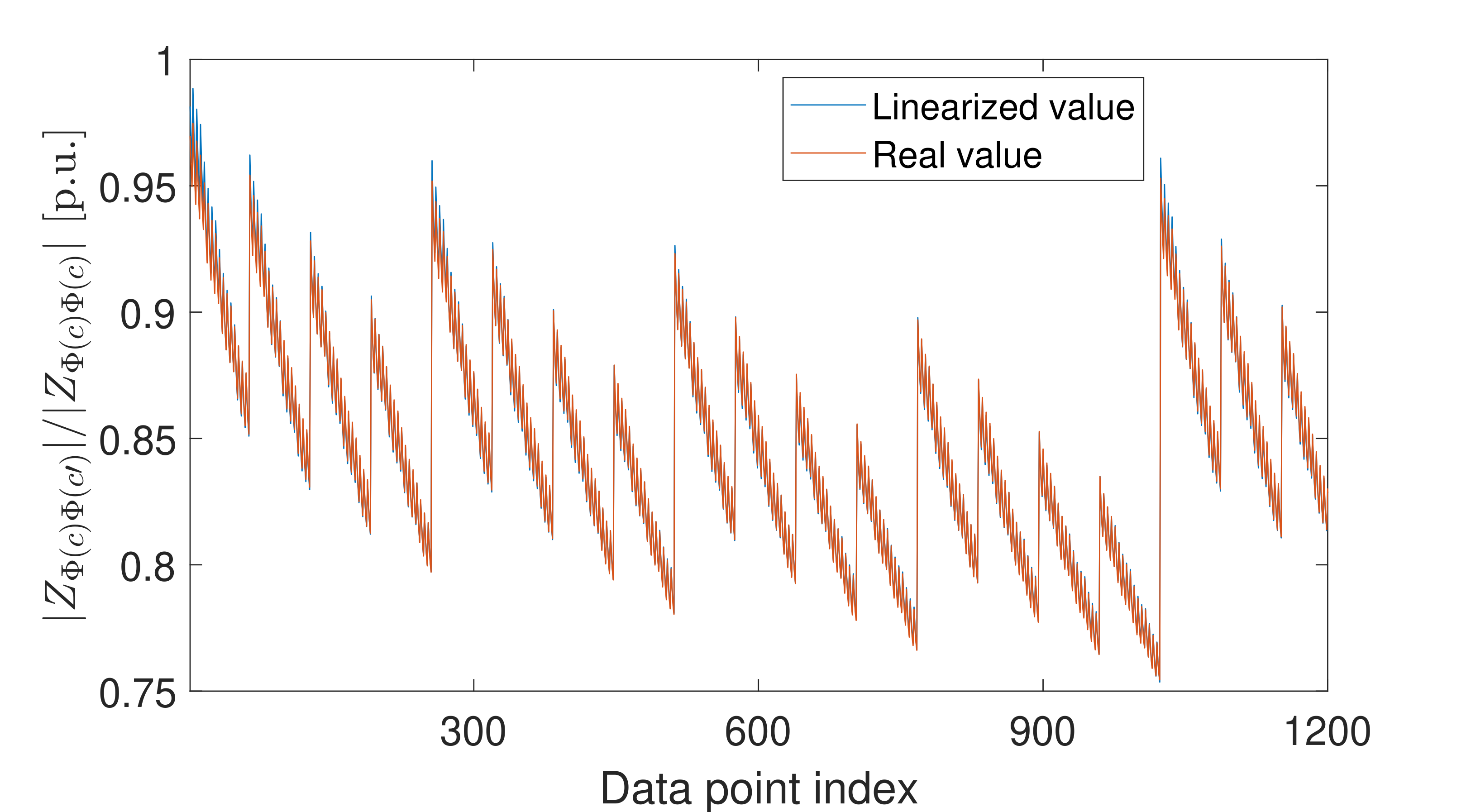}}
    \caption{\label{fig:Linear}Linear regression of the ratio of $Z$ matrix elements.}
\end{figure}

Understandably, $ \mathbf{z}_c^{c'} = {|Z_{\Phi(c)\Phi(c')}|}/{|Z_{\Phi(c)\Phi(c)}|}$ and $\mathbf{z}_c^1 = 1/{|Z_{\Phi(c)\Phi(c)}|}$ include $|\mathcal{G}_c|$ decision variables ($x_{g_c}$) and $|\mathcal{G}_v|$ parameters ($\alpha_{g_v}$). Therefore, their linearized expressions represented by $z_c^{c'}$ and $z_c^1$ respectively, can be formulated as follows:
\begin{align}
    \label{z_linear}
    z_{c}^\delta = \sum_{g_c\in\mathcal{G}_c} k^\delta_{c,g_c} x_{g_c} &+ \sum_{g_v\in\mathcal{G}_v} k^\delta_{c,g_v} \alpha_{g_v} + \sum_{m\in\mathcal{M}} k^\delta_{c,m} \eta_m, \nonumber \\
    & \,\,\forall \delta\in \{c',\,1\}, \,\forall c,c' \in \mathcal{C},\,c'\neq c,
\end{align}
where $k^\delta_{c,g_c}$, $k^\delta_{c,g_v}$ and $k^\delta_{c,m}$ are the linear coefficients. Note that an index $\delta\in \{c',\,1\}$ is introduced to represent $z_c^{c'}$ and $z_c^1$ in a uniform way for simplicity. Due to the original nonlinear nature in $\mathbf{z}_{c}^\delta$, the term $k^\delta_{c,m} \eta_m$ is added in \eqref{z_linear} to describe the interactions between every two units in (V)SGs, i.e., $m\in\mathcal{M} =\{g_1,\,g_2 \mid g_1,\,g_2\in \mathcal{G}\}$. This is defined as follows:
\begin{align}
    \label{eta}
    \eta_m  =
    \begin{cases}
        x_{g_1}x_{g_2},\quad \mathrm{if} \,\,g_1,g_2 \in \mathcal{G}_c\\
        x_{g_1}\alpha_{g_2},\quad \mathrm{if} \,\,g_1\in \mathcal{G}_c\,g_2 \in \mathcal{G}_v \\
        \alpha_{g_1}\alpha_{g_2},\quad \mathrm{if} \,\,g_1,g_2 \in \mathcal{G}_v
    \end{cases},\,\,\forall m \in \mathcal{M}.
\end{align}
Note that since the 2nd order terms lead to accurate enough results, the higher order terms are neglected in order to achieve a balance between the accuracy and computational effort. An example of the regression performance is given in Fig.~\ref{fig:Linear} where a good approximation can be observed for most of the data point. Furthermore, although some errors exist, the corresponding system configurations such as that where only one SG is online, are unlike to occur in practice. Due to the space limitation and the pseudo periodic property, only the first 1200 data points are shown in the figure.

The coefficients $\mathcal{K}^\delta =\{k^\delta_{c,g_c},\,k^\delta_{c,g_v},\,k^\delta_{c,m}\},\,\forall \delta, c,g_c,g_v,m$ are determined by solving the minimization problem $\forall c\in\mathcal{C}$:
\begin{subequations}
\label{DM1}
\begin{align}
    \label{min_linear_coeff}
    & \min_{\mathcal{K}^\delta} \sum_{\omega \in \Omega } \left( z_c^{\delta(\omega)} -\mathbf{z}_c^{\delta(\omega)} \right)^2\\
    \label{z_linear_eva}
    &\left.  z_c^{\delta(\omega)} =  z_c^{\delta} \right\rvert_{x_{g_c}^{(\omega)},\,\alpha_{g_v}^{(\omega)},\,\eta_m^{(\omega)}},
\end{align}
\end{subequations}
where \eqref{z_linear_eva} is calculated according to \eqref{z_linear} and $\omega = \{x_{g_c}^{(\omega)},\,\alpha_{g_v}^{(\omega)},\, \mathbf{z}_c^{\delta(\omega)}\}\in \Omega$ denotes the data set. It is generated by evaluating $\mathbf{z}_c^{\delta(\omega)}$ at each IBG bus in representative system conditions. For the SGs, all the possible generator combinations are considered, whereas for the continuous parameters $\alpha_{g_v}\in [0,\,1]$, there are infinite possible conditions in theory. To obtain the data set with a finite size, the interval is evenly divided into $n_v$ regions, $\forall g_v\in \mathcal{G}_v$, each of which is represented by its mean value. Hence, the total size of $\Omega$ is $2^{|\mathcal{G}_c|}\cdot n_v^{|\mathcal{G}_v|}$.

\subsection{Linearization of Equivalent Power Injection}
Substituting the linearized expression of ${|Z_{\Phi(c)\Phi(c')}|}/{|Z_{\Phi(c)\Phi(c)}|}$ \eqref{z_linear} into \eqref{PQ_1} gives:
\begin{align}
\label{P_linear}
    \hat P_c = P_{c}+  \sum_{c'\in \mathcal{C},c'\neq c}  \left(\sum_{g_c\in\mathcal{G}_c} k^{c'}_{c,g_c} x_{g_c} + \sum_{g_v\in\mathcal{G}_v} k^{c'}_{c,g_v} \alpha_{g_v} \right. & \nonumber \\
     + \left.\sum_{m\in\mathcal{M}} k^{c'}_{c,m} \eta_m \right) P_{c'}&.
\end{align}
Because of the similarity, $\hat Q_c$ is not presented. Since $P_{c'}$ is viewed as a decision variable in the UC problem, the nonlinear terms in \eqref{P_linear} appear in $x_{g_c}P_{c'}$ and $\eta_m P_{c'}$. Considering the expression of $\eta_m$ in \eqref{eta}, only the linearization of the most complex term, i.e., $\mu_{g_1,g_2}^{c'} = x_{g_1}x_{g_2}P_{c'}$ is elaborated in \eqref{bigM} and all the other situations can be dealt with through a similar or even simpler method.
\begin{subequations}
\label{bigM}
\begin{align}
    \label{big_1}
    \mathbf{x}_{g_{12}} &\le x_{g_1} \\
    \mathbf{x}_{g_{12}} &\le x_{g_2} \\
    \label{big_3}
    \mathbf{x}_{g_{12}} &\ge x_{g_1} + x_{g_2}-1 \\
    \label{big_4}
    \mu_{g_1,g_2}^{c'} & \le \mathbf{x}_{g_{12}} P_{c'}^{\mathrm{max}}\\
    \mu_{g_1,g_2}^{c'} & \ge -\mathbf{x}_{g_{12}} P_{c'}^{\mathrm{max}}\\
    \mu_{g_1,g_2}^{c'} & \le P_{c'}+(1-\mathbf{x}_{g_{12}}) P_{c'}^{\mathrm{max}}\\
    \label{big_7}
    \mu_{g_1,g_2}^{c'} & \ge P_{c'}-(1-\mathbf{x}_{g_{12}}) P_{c'}^{\mathrm{max}}
\end{align}
\end{subequations}
\eqref{big_1} - \eqref{big_3} force the relationship of $\mathbf{x}_{g_{12}} = x_{g_1}x_{g_2}$ where $\mathbf{x}_{g_{12}}\in \{0,1\}$ is a binary ancillary variable. \eqref{big_4} - \eqref{big_7} are the exact linearization of $\mu_{g_1,g_2}^{c'} = \mathbf{x}_{g_{12}} P_{c'}$ with $P_{c'}^{\mathrm{max}}$ being the maximum possible value of $P_{c'}$ in magnitude.

\section{Voltage Stability Constrained UC Formulation} \label{sec:4}
In this section, the frequency-constrained SUC model proposed in \cite{9066910} which simultaneously optimizes the system Primary Frequency Response (PFR) and the Synthetic Inertia (SI) from wind turbines is extended to incorporate the voltage stability constraint. With the SOC reformulation of the voltage stability constraint discussed in Section \ref{sec:3}, the whole problem is formulated as MISOCP.

\subsection{Objective Function} \label{sec:3.1}
The objective of the SUC problem is to minimize the expected cost over all nodes in the given scenario tree:
\begin{equation}
    \label{eq:SUC}
    \min \sum_{n\in \mathcal{N}} \pi (n) \left( \sum_{g\in \mathcal{G}}  C_g(n) + \Delta t(n) c^s P^s(n) \right)
\end{equation}
where $\pi(n)$ is the probability of scenario $n\in \mathcal{N}$ and $C_g(n)$ is the operation cost of unit $g\in \mathcal{G}$ in scenario n including startup, no-load and marginal cost; $\Delta t(n)c^sP^s(n)$ represents the cost of the load shedding in scenario n with the three terms being the time step of scenario n, load shedding cost and shed load. 

The scenario tree is built based on user-defined quantiles of the forecasting error distribution to capture the uncertainty associated with the demand and the wind generation. A rolling plan approach is implemented in the SUC model. At each time step, a 24-hour horizon SUC problem is performed with only the decisions of the current node being applied and all the future decisions discarded. At next time step, the scenario tree is updated according to the realizations of the uncertain variables and the process repeats. Due to this rolling planning method, different realizations of nodes in the scenario tree are consistently generated at every hourly timestep according to the forecast error quantile based probability.

The objective function \eqref{eq:SUC} is subjected to a number of constraints. Due to the space limitation, all the conventional UC constraints such as those related to thermal generation units and transmission lines are not listed in the paper. The readers can refer \cite{7115982} for details.

\subsection{AC Power Flow and Power Balance Constraints}
While DC power flow assumption is often made in UC problems of transmission systems, it cannot be applied here since reactive power is of concern in the voltage stability assessment. Instead, AC power flow and its SOCP relaxation recently developed in \cite{Kocuk2016} are adapted as follows. For each line $ij\, \in \mathcal{L} $, define the ancillary variables:
\begin{subequations}
\label{Def_CS}
    \begin{align}
        c_{ij} &= |V_i||V_j|\cos(\theta_i- \theta_j) \\
        s_{ij} &= -|V_i||V_j|\sin(\theta_i- \theta_j) ,
    \end{align}
\end{subequations}
where $|V_i|\angle{\theta_i}$ is the voltage at Bus i. As a result, the AC power flow constraints in SOC form can be formulated as: 
\begin{align}
    &P_i^G-P_i^D = G_{ii}c_{ii} + \sum_{j\in L(i)} P_{ij},\,\,\;\,\forall i \in \mathcal{B} \label{AC_P} \\ 
    &Q_i^G-Q_i^D = - B_{ii}c_{ii} + \sum_{j\in L(i)} Q_{ij},\,\,\;\,\forall i \in \mathcal{B} \label{AC_Q} \\
    &V_{\mathrm{min},i}^2\le c_{ii} \le V_{\mathrm{max},i}^2,\,\,\;\,\forall i \in \mathcal{B} \label{AC_V} \\
    & c_{ij} = c_{ji}, \,\,s_{ij} = -s_{ji} ,\,\,\;\,\forall ij \in \mathcal{L} \label{AC_cs}\\
    & c_{ij}^2+s_{ij}^2\le c_{ii}c_{jj},\,\,\;\,\forall ij \in  \mathcal{L}.  \label{AC_soc}
\end{align}
\eqref{AC_P} and \eqref{AC_Q} are the constraints of active and reactive power balance at Bus i with the superscript $G,D$ representing generation and demand and $G_{ii}+\mathrm{j} B_{ii}$ is the diagonal element of nodal admittance matrix $Y$; \eqref{AC_V} is the voltage magnitude constraints at all buses; \eqref{AC_cs} and \eqref{AC_soc} are the SOC relaxation of \eqref{Def_CS}.

\subsection{Voltage Stability and IBG Capacity Constraints}
The voltage stability constraints in SOC form developed in Section \ref{sec:3} are summarized as follows:
\begin{subequations}
\label{VS_SOC}
\begin{align}
    & \hat{P}_c^2+\hat{Q}_c^2\le { \left( \hat{Q}_c+ {\Gamma_c} \right) ^2} ,\,\,\;\,\forall c \in  \mathcal{C}\\
    & \hat P_c = P_{c}+  \sum_{c'\in \mathcal{C},c'\neq c}  \left(\sum_{g_c\in\mathcal{G}_c} k^{c'}_{c,g_c} x_{g_c} + \sum_{g_v\in\mathcal{G}_v} k^{c'}_{c,g_v} \alpha_{g_v} \right. \nonumber \\
    & \qquad \qquad \qquad \qquad + \left.\sum_{m\in\mathcal{M}} k^{c'}_{c,m} \eta_m \right) P_{c'},\,\,\;\,\forall c \in  \mathcal{C} \label{P_linear}\\
    & \hat Q_c = Q_{c}+  \sum_{c'\in \mathcal{C},c'\neq c}  \left(\sum_{g_c\in\mathcal{G}_c} k^{c'}_{c,g_c} x_{g_c} + \sum_{g_v\in\mathcal{G}_v} k^{c'}_{c,g_v} \alpha_{g_v} \right. \nonumber \\
    & \qquad \qquad \qquad \qquad + \left.\sum_{m\in\mathcal{M}} k^{c'}_{c,m} \eta_m \right) Q_{c'},\,\,\;\,\forall c \in  \mathcal{C} \label{Q_linear}\\
    & \Gamma_c = 2 \left(\sum_{g_c\in\mathcal{G}_c} k^{1}_{c,g_c} x_{g_c} + \sum_{g_v\in\mathcal{G}_v} k^{1}_{c,g_v} \alpha_{g_v} + \sum_{m\in\mathcal{M}} k^{1}_{c,m} \eta_m \right) \nonumber \\
    &\qquad \qquad \qquad \qquad \qquad \qquad \qquad \qquad \qquad \;\; \forall c \in  \mathcal{C}. \label{gamma_linear}
\end{align}
\end{subequations}
The nonlinearity in \eqref{P_linear}-\eqref{gamma_linear} can be effectively linearized through the method in \eqref{bigM}.
Note that only the voltage stability at the buses of grid-following IBGs is considered since they are the most vulnerable buses in high PE-penetrated system due to the large electrical distance from rest of the system and the low system strength at the interconnection point. 

Based on the analysis in Section \ref{sec:2}, it is clear that increasing the reactive power injection from grid-following IBGs improves the voltage stability. However, this improvement is limited due to the IBGs' capacity constraints in steady state:
\begin{equation}
     {P}_c^2+{Q}_c^2\le S_{c,\mathrm{max}}^2 ,\,\,\;\,\forall c \in  \mathcal{C}.
\end{equation}
Although temporary overloading of $1.2 - 1.5\mathrm{p.u.}$ is allowable for IBGs \cite{9329077}, it is not feasible during steady state operations. Therefore, $S_{c,\mathrm{max}} = 1\,\mathrm{p.u.}$ is selected in this paper. 

\subsection{Frequency Constraints} \label{sec:4.4}
Since both the frequency and the voltage stability constraints are influenced by the SGs and IBGs, it is of interest to investigate their combined effects on system operation. Therefore, the frequency security constraints are elaborated here. A novel wind turbine SI control scheme is proposed in \cite{9066910}, which eliminates the secondary frequency dip due to WT overproduction and allows the WT dynamics to be analytically integrated into the system frequency dynamics. The frequency nadir constraint is formulated as:
\begin{equation}
\label{nadir_c}
    HR\ge \frac{\Delta P_L^2T_d}{4\Delta f_\mathrm{lim}}-\frac{\Delta P_L T_d }{4} \left(D- \sum_{j\in \mathcal{F}} \gamma_j H_{s_j}^2 \right),
\end{equation}
where $H$ is the sum of conventional inertia $H_c$ and SI of all the wind farms $\sum_{j} H_{s_j}$ with $j\in \mathcal{F}$ being the set of wind farms and $R,\,T_d$ represent the system Primary Frequency Response (PFR) and its delivery time; $\Delta P_L,\, \Delta f_{\mathrm{lim}}$ and $D$ denote system disturbance, limit of frequency deviation and system damping. The last term can be interpreted as: SI provision from each wind farm introduces a negative damping to the system proportional to $H_{s_j}$ with the coefficient being $\gamma_j$. In order to fit the MISOCP-based SUC model, the frequency nadir constraint \eqref{nadir_c} is dealt with through SOC reformulation similar to the method in \cite{9475967} instead of piece-wise linearization in \cite{9066910}. Rewritten \eqref{nadir_c} as follows:
\begin{equation}
\label{nadir_soc}
    HR\ge \underbrace{\frac{\Delta P_L^2T_d}{4\Delta f_\mathrm{lim}}-\frac{\Delta P_L T_d D }{4}}_{x_1^2} + \frac{\Delta P_L T_d \sum_{j\in \mathcal{F}} \gamma_j H_{s_j}^2}{4} ,
\end{equation}
which is a standard form of a multi-dimensional rotated second-order cone, with the decision variables being $H,\,R,\,x_1$ and $H_{s_j}$. Being a constant, the ancillary variable $x_1$ is defined for the sole purpose of SOC reformulation. Additionally, since in real power systems, the relationship $\Delta P_L/\Delta f_\mathrm{lim}>D$ always holds, $x_1$ is real and the SOC nadir constraint \eqref{nadir_soc} is well defined.

\section{Case Studies}  \label{sec:5}
Case studies on the IEEE 30-bus systems are carried out to illustrate the validity of the proposed model and the impacts of the voltage stability constraint on system operation as well as its interaction with system frequency constraints. A stability margin of $5\%$ is included to insure the conservativeness of the voltage stability. The network data of the systems can be obtained in \cite{Data_system}. The MISOCP-base UC problem is solved by FICO Xpress on a PC with Intel(R) Core(TM) i7-7820X CPU @ 3.60GHz and RAM of 64 GB. 

\begin{figure}[!t] 
	\centering
	\vspace{-0.4cm}
	\scalebox{0.25}{\includegraphics[]{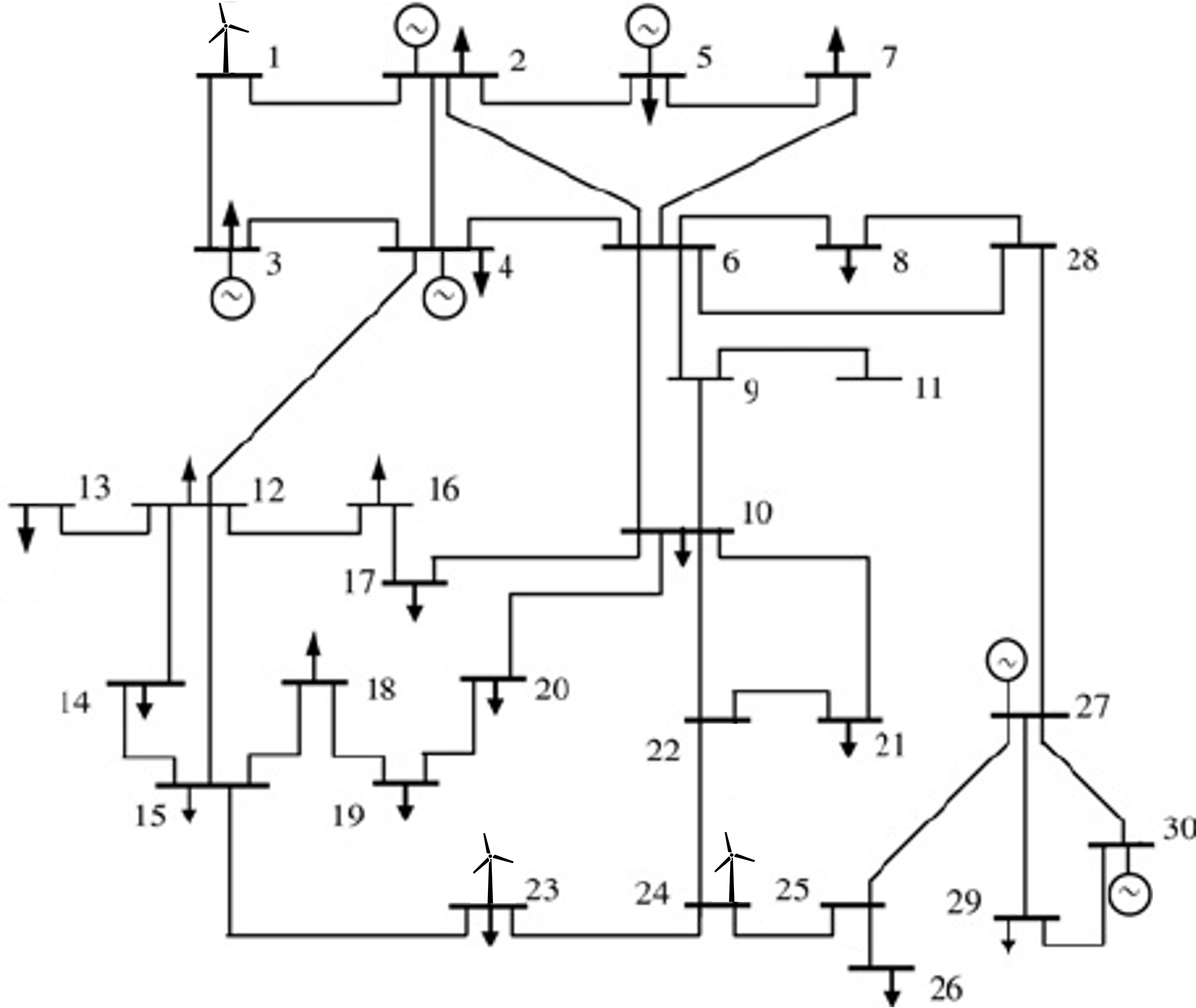}}
	\caption{Modified IEEE-30 bus system.}
	\label{fig:ieee-30}
\end{figure}

The system is modified by adding wind generation at Bus 1, 23 and 24 to model the IBG penetration and voltage stability issues with the first being grid-forming and the rest grid-following, as shown in Fig.~\ref{fig:ieee-30}. The system parameters are set as follows: base MVA $S_B=100\,\mathrm{MVA}$ load demand $P_D \in [160,410] \,\mathrm{MW}$, damping $D = 0.5\% P_D / 1\,\mathrm{Hz}$, FR delivery time $T_d = 10\,\mathrm{s}$ and maximum power loss $\Delta P_L = 50\,\mathrm{MW}$. The frequency limits of nadir, steady state and RoCoF set by National Grid are: $\Delta f_\mathrm{lim} = 0.8\,\mathrm{Hz}$, $\Delta f_\mathrm{lim}^\mathrm{ss} = 0.5\,\mathrm{Hz}$ and $\Delta \dot f_\mathrm{lim} = 0.5\,\mathrm{Hz/s}$.

\subsection{Impact of Wind Penetration on Voltage stability}  \label{sec:5.1}
\begin{figure}[!t]
    \centering
	\scalebox{1.14}{\includegraphics[]{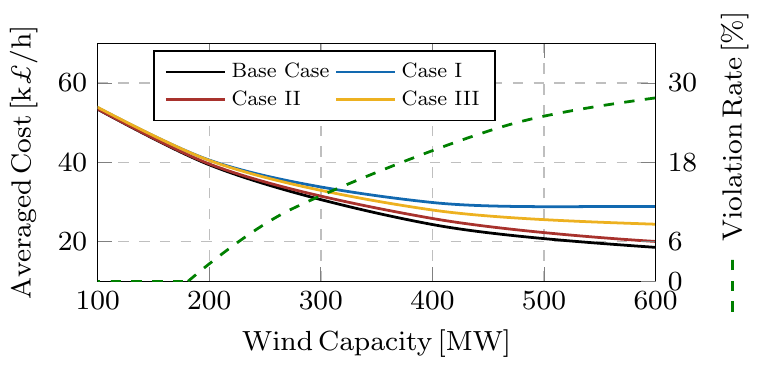}}
    \caption{\label{fig:Cost_Wind}System operation cost with various installed wind capacity.}
\end{figure}

This section investigates the influence of system wind penetration on the voltage stability at grid-following IBG buses. Three different cases are considered in the SUC problem.
\begin{itemize}
    \item Base Case: voltage stability constraints are neglected
    \item Case I: voltage stability constraints are included without allowing reactive power from IBGs
    \item Case II: voltage stability constraints are included with reactive power from IBGs allowed
    \item Case III: voltage stability constraints are included with reactive power support from $50\%$ of the IBGs
\end{itemize}
The averaged system operation cost over one year operation in different cases is depicted in Fig.~\ref{fig:Cost_Wind} against the total installed wind capacity at Bus 23 and 24. It is understandable that in Base Case where voltage stability is not considered, the averaged operational cost decreases as more wind generation is installed in the system. However, in this case, the voltage stability violation (green dashed curve in Fig.~\ref{fig:Cost_Wind}) starts to appear when the wind capacity comes close to $200\,\mathrm{MW}$ and keep increasing as increased wind capacity. 

On the other hand, once the voltage stability constraint is applied in Case I,
from the blue curve, it can be observed that significant cost increase incurs to maintain such constraint. In particular, after the wind capacity reaching $400\,\mathrm{MW}$, the system operation cost remains at around $30\,\mathrm{k\pounds/h}$ and ceases further reduction regardless of the wind capacity. This is because the voltage stability constraints at the IBG buses prevent more active power being injected to the gird. As a result, the installed wind power cannot be utilized and hence the system operation cost barely changes.
In order to make use of more wind power while maintaining the voltage stability, reactive power is allowed to be injected to the grid by IBGs (Case II). The averaged system operation cost in this case almost aligns with that in Base Case, indicating the effectiveness of the reactive power on improving the voltage stability. This improvement becomes less obvious if only a portion ($50\%$) of the IBGs can provide reactive support, as indicated by the yellow curve. 

In summary, it is clear that voltage stability in high PE-penetrated system should be considered during the scheduling process and appropriate reactive power injections from IBGs significantly reduce the cost increment of maintaining voltage stability.

\subsection{Impact of System Strength on Voltage stability} \label{sec:5.2}
\begin{figure}[!t]
    \centering
	\scalebox{1.14}{\includegraphics[]{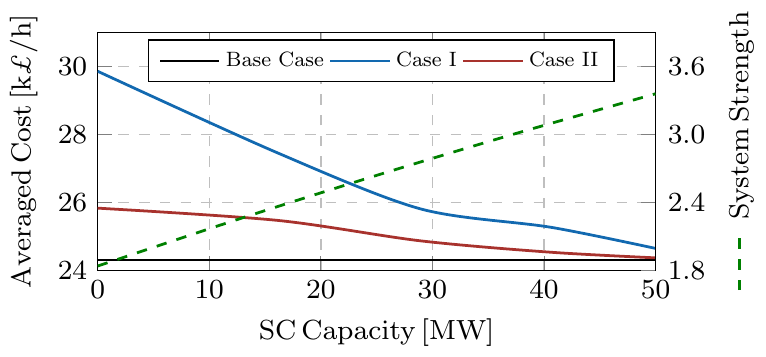}}
    \caption{\label{fig:Cost_Zii}System operation cost with different system strength.}
\end{figure}
Another factor that influences voltage stability is the system strength at IBG buses. It is closely related to the electrical distance from the IBGs to the rest of the system. To assess this impact, the SUC problem is performed with the system strength at Bus 23 and 24 varying at the same time. This is achieved by adding a Synchronous Condenser (SC) with different capacity at Bus 22. Notably, the wind capacity in this subsection is selected to be $400\,\mathrm{MW}$. The result is plotted in Fig.~\ref{fig:Cost_Zii} with the system strength representing the average of Bus 23 and 24, i.e., $\frac{1}{2}\left(\frac{1}{Z_{23,23}}+\frac{1}{Z_{24,24}}\right)$ among the entire time horizon. This averaged system strength increases linearly with the SC capacity at Bus 22. In Base Case where voltage stability constraints are not considered, the system operation cost is almost irrelevant with the SC capacity. 

On the contrary, the average system operation cost decreases as the SC capacity increases in Case I, with an approximately linear rate. This is due to the fact that more active power can be transferred from IBGs to the grid with higher system strength without violating the voltage stability constraints. It suggests that the voltage stability issues become less significant as the grid gets stronger. Since large scale PE-based generation units are generally located electrically distant to load centers, system operator can also mitigate the voltage stability problem by increasing the system strength, such as installing SC and transmission network enhancement.

Moreover, if reactive power injection is allowed (red curve), a similar trend can be spotted. However, the decline rate of system operation cost is less significant compared with that in Case I. Therefore, it may not be economic benefit to further reduce the system operation cost by installing additional SC, if the reactive power support from IBGs is available and the optimal strategy would require further investigation.

\subsection{Assessment of IBGs' Interaction}    \label{sec:5.3}
As analyzed in Section \ref{sec:2}, the IBGs electrically close, interact with each other and influence their voltage stability. This section reveals how IBGs' interaction impacts the voltage stability and system operation. First, the SUC problem is performed without considering the interaction of the IBGs. The results are then tested against the real voltage stability condition considering the interaction. A violation rate of $13.7\%$ is found, thus highlighting the necessity of considering the IBGs' interaction when assessing the voltage stability.

\begin{figure}[!t]
    \centering
	\scalebox{1.2}{\includegraphics[]{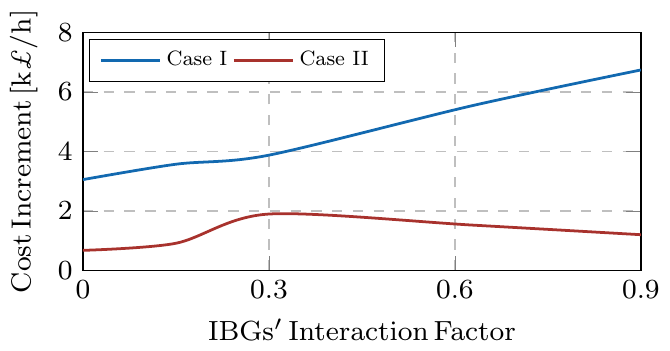}}
    \caption{\label{fig:Cost_Zr}System operation cost with different IBGs' interaction factors.}
\end{figure}

Furthermore, to investigate the influence of the electrical distance between the IBGs, the SUC is conducted with different IBGs' interaction factor $\xi$, representing the averaged ratio of referred power injection from other IBGs to the self power injection:
\begin{equation}
    \xi = \frac{1}{|\mathcal{C}|}\sum_{c\in \mathcal{C}}\frac{\hat P_c-P_c}{P_c}.
\end{equation}
Note that $\xi=0.63$ in the original system (Section \ref{sec:5.1}) and the wind capacity in this section is $400\,\mathrm{MW}$. In Base Case, the IBGs' interaction factor only influences the operating condition through power flow distribution and the system operation cost does not change significantly, thus not being shown. Instead, the system operation cost increments in Case I and II compared with Base Case are plotted in Fig.~\ref{fig:Cost_Zr}.

When the voltage stability constraints are implemented (Case I), a clear increasing trend is presented in the operational cost as the IBGs' interaction factor grows. This is because when the IBGs becomes electrically closer to each other, the equivalent active power injection from each IBG increases as defined in \eqref{PQ_1} and to ensure the voltage stability, the actual active power output from wind generation has to be reduced, thus leading to a cost increment. Finally, with the reactive power support from IBGs (Case II), the system operation cost increases first and then decreases as the IBGs' interaction becomes stronger, compared with that in Base Case. During this process, on one hand, the increased equivalent active power injection deteriorates the voltage stability; On the other hand, the increased equivalent reactive power injection mitigate it. The latter dominates the overall effect after around $\xi=0.3$ where a maximum increment compared to Base Case is spotted. 

\subsection{Value of Reactive Power Optimization}
\begin{figure}[!t]
    \centering
	\scalebox{1.2}{\includegraphics[]{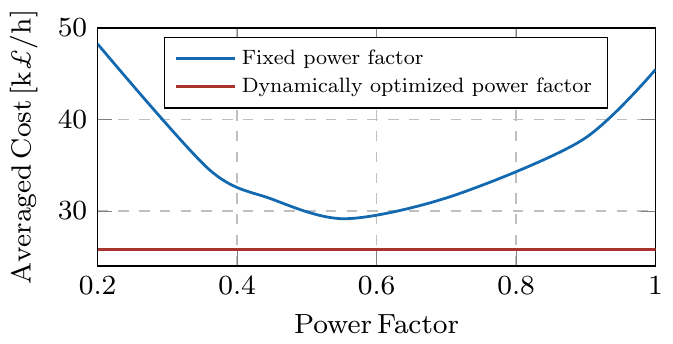}}
    \caption{\label{fig:PF}System operation cost with alternative reactive power supply methods.}
\end{figure}
Based on the analysis in the previous three subsections, it can be concluded  that reactive power injection is very effective in improving the voltage stability. It has been considered by some grid standards to require IBGs to supply reactive power with a fixed pre-defined power factor, which is, however, not optimal given the varying system conditions. To demonstrate that, this subsection compares the fixed power factor approach with the proposed one where the reactive power injection of IBGs in each bus is optimized at each time step. As illustrated in Fig.~\ref{fig:PF}, as the power factor reduces from $1$, the increased reactive power helps to improve the voltage stability and hence the cost is decreased until $0.55$, where the operation cost is ``minimized''. After that, the reduced power factor continues to decrease the IBGs active power output, resulting in a cost growth. Nevertheless, it is not straightforward to obtain the best pre-defined power factors for all the IBGs, which is a combinatorial optimization problem. In addition, the best fixed power factor still leads to a higher cost ($29.2\,\mathrm{k\pounds/h}$) than that of the proposed method ($25.8\,\mathrm{k\pounds/h}$) since the power factor of each IBG in the proposed method is optimized at each time step. Note that the red curve in the figure is depicted for comparison and is irrelevant to the horizontal axis.

\subsection{Interaction of Voltage Stability with Frequency Constraints and SI Provision}
\begin{figure}[!b]
    \centering
	\scalebox{1.2}{\includegraphics[]{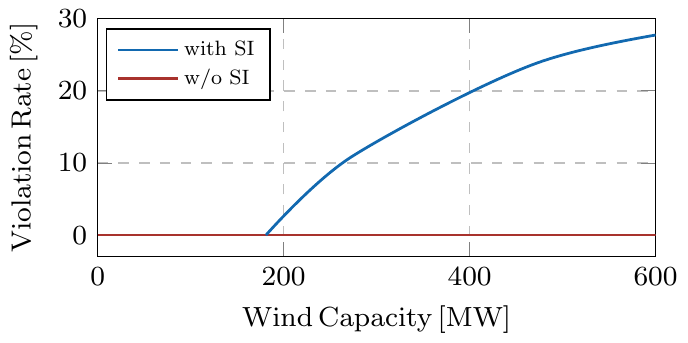}}
    \caption{\label{fig:SI}Voltage stability violation rate with and without SI provision.}
\end{figure}
Since both the voltage stability and frequency security depend on the status of synchronous generators and the amount of available wind power, it is worthwhile to investigate the interaction of voltage stability with the frequency constraints and the SI provision. As illustrated in Fig.~\ref{fig:SI}, the violation rate of voltage stability constraints when they are not incorporated into the SUC model is zero if no SI from wind turbines is available. However, after including the SI, the violation rate starts to appear as the wind capacity comes close to $200\, \mathrm{MW}$ and then increase gradually. This is due to the fact that with the SI provision from wind generation units, less SGs are required to provide inertia for frequency constraint maintenance. As a consequence, more wind power can be utilized and the system strength at the IBG buses is reduced, both of which make the voltage stability problem severer. It should be noted that the SI provision can be replaced by any other frequency support methods from IBGs with the analysis and conclusion still holding. 

\section{Conclusions}    \label{sec:6}
This paper presents a voltage stability constrained UC problem where the system operation cost is minimized while maintaining the static voltage stability of IBG buses by dynamically optimizing the active and reactive power injection from IBGs. The interaction of different IBGs is explicitly considered when formulating the voltage stability constraints, which are further reformulated into SOC form. Together with the SOC relaxation of the AC power flow and frequency stability constraints, the overall problem is represented as MISOCP. Case studies demonstrate that including voltage stability constraints increases the system operation cost especially with high PE penetration and this increment can be considerably reduced when reactive power support from IBGs is allowed. Moreover, increasing the system strengths at IBG buses and the electrical distance between IBGs improves the voltage stability. Dynamically optimizing the reactive power supply from IBGs is also illustrated to be superior than fixed power factor provision in terms of more efficient system operation. The voltage stability problem would become dominating in high PE-penetrated system, only after the frequency challenge is resolved by SI provision.

\vspace{-0.25cm}
\bibliographystyle{IEEEtran}
\bibliography{bibliography}
\end{document}


\begin{varwidth}{\linewidth}

\begin{tikzpicture}
\begin{axis}[
    scaled ticks=false,
    tick label style={/pgf/number format/fixed},
    colormap name=viridis,
    width=7.25cm,
    height=4cm,
    xlabel={$\mathrm{Wind\,Capacity\,[MW]}$},
    ylabel={$\mathrm{Violation\, Rate\,[\%]}$},
    xmin=0, xmax=600,
    ymin=-3, ymax=30,
    xmajorgrids=true,
    ymajorgrids=true,
    legend style={at={(axis cs:9, 28.67)},anchor=north west,nodes={scale=0.75, transform shape}, legend columns=1},
    legend cell align={left},
    grid style=dashed,
]
\footnotesize

\addplot[
    thick,
    color=pBlue,
    ]
    table {data/SI/data3.txt};        
    \addlegendentry{\footnotesize with SI} 
\addplot[
    thick,
    color=pRed,
    ]
    table {data/SI/data2.txt};        
    \addlegendentry{\footnotesize w/o SI} 
\addplot[
    smooth,
    thick,
    color=pBlue,
    ]
    table {data/SI/data1.txt};    

\end{axis}

\end{tikzpicture} 

\end{varwidth}